\documentclass[conference]{IEEEtran}
\IEEEoverridecommandlockouts

\usepackage{cite}
\usepackage{amsmath,amssymb,amsfonts}
\usepackage{algorithmic}
\usepackage{graphicx}
\usepackage{textcomp}
\usepackage{xcolor}
\usepackage{amsmath,graphicx}
\usepackage{multirow}
\usepackage{booktabs}
\usepackage{amsfonts}
\usepackage{multirow}
\usepackage{bbding}
\usepackage{cite}
\usepackage[urlcolor=blue]{hyperref}
\usepackage{mathrsfs}
\usepackage{threeparttable}
\usepackage{adjustbox}
\usepackage{float}
\usepackage{caption}
\usepackage{graphicx}
\usepackage{float} 
\usepackage{graphicx}
\usepackage{float}
\usepackage{subfig}

\def\BibTeX{{\rm B\kern-.05em{\sc i\kern-.025em b}\kern-.08em
    T\kern-.1667em\lower.7ex\hbox{E}\kern-.125emX}}
\begin{document}

\title{Adversarial Speaker Disentanglement Using Unannotated External Data for Self-supervised Representation-based Voice Conversion\\
}
\makeatletter
\newcommand{\linebreakand}{%
  \end{@IEEEauthorhalign}
  \hfill\mbox{}\par
  \mbox{}\hfill\begin{@IEEEauthorhalign}
}
\makeatother

\author{
\IEEEauthorblockN{1\textsuperscript{st} Xintao Zhao*}
\IEEEauthorblockA{\textit{Shenzhen International Graduate School} \\
\textit{Tsinghua University}\\
Shenzhen, China \\
zxt20@mails.tsinghua.edu.cn}
\and
\IEEEauthorblockN{2\textsuperscript{nd} Shuai Wang†}
\IEEEauthorblockA{\textit{Lightspeed \& Quantum Studios} \\
\textit{Tencent Inc}\\
Shenzhen, China \\
wsstriving@gmail.com}
\and
\IEEEauthorblockN{3\textsuperscript{rd} Yang Chao}
\IEEEauthorblockA{\textit{Lightspeed \& Quantum Studios} \\
\textit{Tencent Inc}\\
Shenzhen, China \\
youngchao@tencent.com}
\linebreakand
\IEEEauthorblockN{4\textsuperscript{th} Zhiyong Wu†}
\IEEEauthorblockA{\textit{Shenzhen International Graduate School} \\
\textit{Tsinghua University}\\
Shenzhen, China \\
zywu@se.cuhk.edu.hk}
\and
\IEEEauthorblockN{5\textsuperscript{th} Helen Meng}
\IEEEauthorblockA{\textit{Department of Systems Engineering and Engineering Management} \\
\textit{The Chinese University of Hong Kong}\\
Hong Kong SAR, China \\
hmmeng@se.cuhk.edu.hk}
}

\maketitle
\renewcommand{\thefootnote}{\fnsymbol{footnote}} 
\footnotetext[1]{Work done during the internship at Tencent Lightspeed \& Quantum Studios}
\footnotetext[2]{Corresponding author}
\begin{abstract}
Nowadays, recognition-synthesis-based methods have been quite popular with voice conversion (VC). By introducing 
linguistics features with good disentangling characters 
extracted from an automatic speech recognition (ASR) model, the VC performance achieved considerable breakthroughs. 
Recently, self-supervised learning (SSL) methods trained with a large-scale unannotated speech corpus have been applied to downstream tasks focusing on the content information, which is suitable for VC tasks. 
However, a huge amount of speaker information in SSL representations degrades timbre similarity and the quality of converted speech significantly. 
To address this problem,
we proposed a high-similarity any-to-one voice conversion method with the input of SSL representations. We 
incorporated adversarial training mechanisms in the synthesis module using external unannotated corpora. 
Two auxiliary discriminators were trained to distinguish whether a sequence of mel-spectrograms has been converted by the acoustic model and whether a sequence of content embeddings contains speaker information from external corpora. Experimental results show that our proposed method achieves comparable similarity and higher naturalness than the supervised method, which needs a huge amount of annotated corpora for training
and is applicable to improve similarity for VC methods with other SSL representations as input.

\label{sec:abs}
\end{abstract}


\section{Introduction}
Voice conversion (VC) aims to modify the timbre information of one utterance from the source speaker to make it sound like the target speaker. Commonly, the content and prosody information should be kept during the conversion. Thus, the core problem for VC is to achieve an effective disentanglement of the speaker and non-speaker information. Conventional VC methods collect paired audio corpus with the same content information, which is expensive in practice. Non-parallel VC is a more challenging but practical task that only needs training utterances recorded by the target speaker\cite{wang2021towards,wang2021adversarially,choi2021neural}. 
Recognition-synthesis-based VC is one representative solution of the non-parallel VC\cite{sun2016phonetic,zhao2022disentangling,liu2021fastsvc,wang2022one}. By utilizing an automatic speech recognition (ASR) model pretrained with a massive amount of data, robust linguistic representations containing content-related information can be extracted.
Compared to the unsupervised VC method\cite{qian2019autovc}, the recognition-synthesis-based method is much more robust due to the prior knowledge from the ASR model and its pretraining data. In \cite{zhao2022disentangling}, the ASR model trained with Connectionist Temporal Classification loss (CTC-ASR) can provide representations with good disentangling characteristics, which lead to high similarity in VC performance. As shown in Fig.\ref{fig:Rec_Syn_model}(a), an ordinary recognition-synthesis structure extracts bottleneck features from a hidden layer of the ASR model and then feeds it into the synthesis model.
However, training ASR models requires large-scale annotated corpora and even frame-level-aligned transcription, which inevitably leads to an expensive and time-consuming data collection process. 

\begin{figure}[t]
  \centering
  \includegraphics[width=1.0\linewidth]{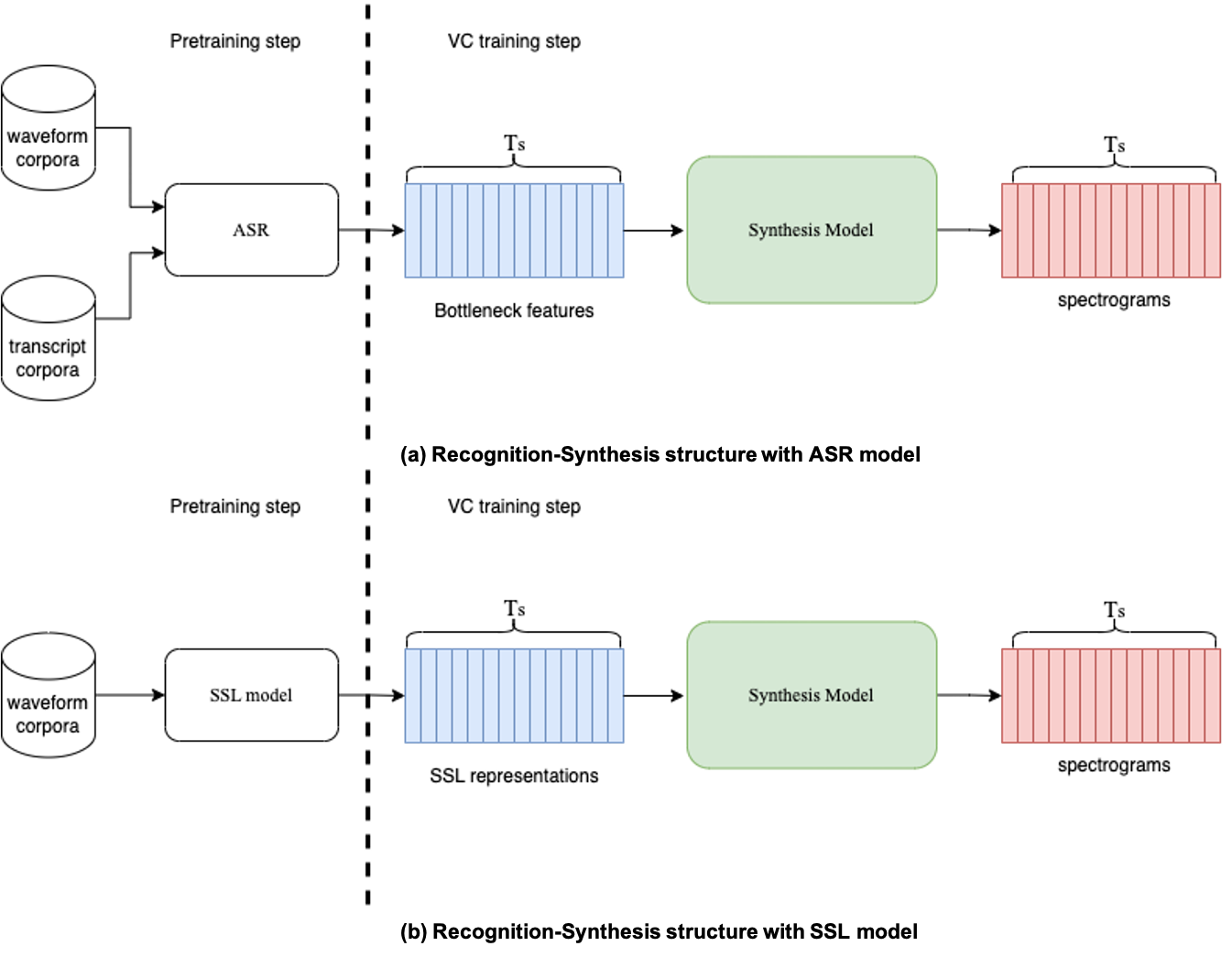}
  \caption{Recognition-synthesis based model architecture.$T_s$ means the number of frames in time axis.}
  \label{fig:Rec_Syn_model}
\end{figure}

Over the recent years, self-supervised learning (SSL) trained on a large-scale unannotated speech corpus\cite{baevski2020wav2vec,hsu2021hubert,van2022comparison,qian2022contentvec} has been applied to downstream tasks focusing on the content information, which also shows considerable potential for VC tasks. Same as Fig.\ref{fig:Rec_Syn_model}(b), we can replace ASR model with SSL model. In addition, the experiments in \cite{wang2021towards} show that the SSL representations could preserve the prosody information contained in the source audio to a good extent. 

However, few existing SSL representations can achieve a reasonable disentanglement between speaker and content information. The leakage of residual speaker information may degrade the VC performance in similarity and quality.

To make the SSL pertaining process focus more on the content modeling, \cite{gat2022speaker} proposed a method to normalize speaker characteristics via negating the gradients of the SSL upstream model concerning the speaker identification task. \cite{van2022comparison} proposed Hubert Soft units as a middle-ground between continuous raw features and discrete labels, which could be considered a speaker-independent representation. In \cite{qian2022contentvec}, Qian et al. proposed ContentVec, an SSL teacher-student framework combined with speaker disentanglement techniques. Though these SSL methods successfully reduced speaker information contained in SSL representation to some extent, the degradation of VC performance caused by residual speaker information still exists.

In this work, we proposed a recognition-synthesis-based SSL VC method that can further reduce residual speaker information contained in SSL representations via adversarial training techniques and external unannotated corpora. There are two auxiliary discriminators in the proposed method. 
A conversion discriminator was trained to distinguish whether a sequence of Mel-spectrograms has been converted by the acoustic model, whose source speech comes from the external corpora. 
An embedding discriminator was trained to distinguish whether a sequence of content embedding is extracted from the external corpora, which contain different speaker information from the target speaker. The proposed method reduces the residual speaker information leaked from SSL representations and significantly improves timbre similarity while retaining the good prosodic characteristics of SSL representations. 
The contributions of this paper are as follows:
\begin{itemize}
  \item We proposed a new adversarial training strategy that could further reduce residual speaker information 
  using external unannotated corpora.
  \item We proposed a feed-forward transformer-based any-to-one VC acoustic model with two auxiliary discriminators, which could generate Mel-spectrograms with high similarity as well as quality.
  \item Experimental results show that the proposed method achieves comparable similarity and higher naturalness than the supervised CTC-ASR-based method and applies to VC methods using other SSL representations as input.
\end{itemize}



\label{sec:intro}

\section{Method}
This section describes our proposed method, which consists of three components, a pretrained SSL model, an acoustic model with a group of discriminators, and a pretrained HiFi-GAN vocoder.
We extract representations from the SSL model and feed them into the acoustic model, as shown in Fig.\ref{fig:proposed_img}, to generate Mel-spectrograms. 
Finally, we feed the generated Mel-spectrograms into the vocoder to get a high-quality waveform. 
The acoustic model adopts the encoder-decoder structure as the backbone, with three different discriminators to determine
(i) Whether a sequence of Mel-spectrograms is extracted from the ground-truth audio or reconstructed by the acoustic model.
(ii) Whether a sequence of Mel-spectrograms has been converted by the acoustic model, whose source speech comes from the external corpora.
(iii) Whether a sequence of content embeddings are extracted from the external corpora, which contain speaker information different from the target speaker.
\begin{figure*}[t]
  \centering
  \includegraphics[width=1.0\linewidth]{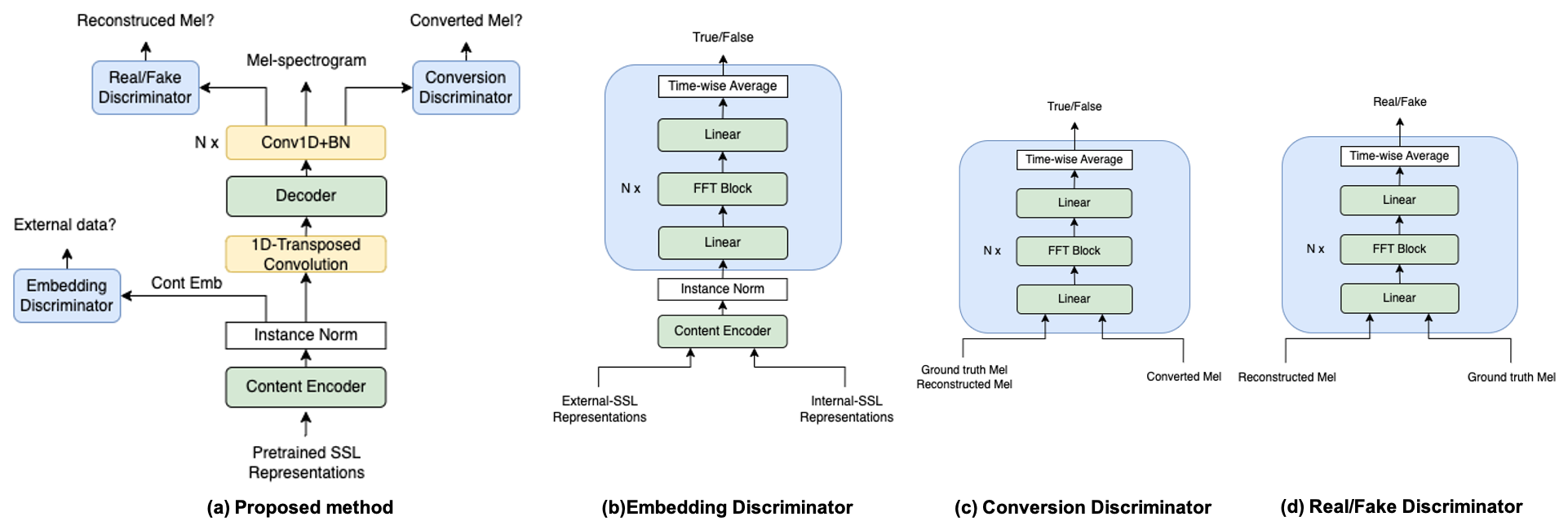}
  \caption{The proposed acoustic model, FFT Block is the feed forward transformer structure proposed in \cite{ren2019fastspeech}. (a). The overall structure. (b). The Embedding Discriminator distinguishes whether an input content embedding is extracted from external corpora. (c) The Conversion Discriminator distinguishes whether an input Mel has been converted by the acoustic model. (d) The Real/Fake Discriminator distinguishes whether an input Mel is reconstructed by the acoustic model.}
  \label{fig:proposed_img}
\end{figure*}

\subsection{Self supervised model}

As for the pretrained SSL model, we select three kinds of SSL frameworks that are adapted from the Hubert training paradigm as the candidates:

\textbf{Hubert Raw feature} We extracted the Hubert Raw features from an intermediate layer in a pretrained Hubert model. Following \cite{van2022comparison}, we used the output of the seventh transformer layer, whose resulting acoustic units perform well on phone discrimination tests\cite{van2021analyzing}.

\textbf{Hubert Soft unit} We use the Hubert Soft unit proposed in \cite{van2022comparison} as one kind of input SSL representation, which is trained by predicting clustered labels via a linear-layer-based soft encoder. Hubert Soft unit has much fewer dimensions than the Hubert Raw feature and brings notable improvement in similarity as well as naturalness.

\textbf{ContentVec:} ContentVec\cite{qian2022contentvec} is a new SSL framework that can achieve speaker disentanglement without severe loss of content. They introduced an any-to-one VC into a teacher model to remove the residual speaker information contained in clustered labels and adopt the random perturb algorithm proposed in \cite{choi2021neural} for speaker disentanglement. By combining the Hubert-based teacher-student framework with speaker disentanglement techniques, good disentangling characteristics are retained in ContentVec representations. 

\subsection{Acoustic model}

The proposed acoustic model follows an encoder-decoder structure. The input SSL representations are passed through a content encoder followed by instance normalization (IN)\cite{chou2019one}. The outputs of the content encoder are then fed into a 1D-transposed convolution and a decoder. Both content encoder and decoder are based on the feed-forward transformer (FFT) structure proposed in\cite{ren2019fastspeech} followed by a linear projection layer. The generated Mel-spectrograms from the decoder are then fed into a HiFi-Gan-based vocoder\cite{kong2020hifi} to generate the high-quality waveform.

\subsection{Discriminator group}

\textbf{Real/Fake discriminator} The real/fake discriminator aims to distinguish whether a sequence of Mel-spectrograms is ground truth or reconstructed by the acoustic model. Note that the source speech of reconstructed Mel comes from the target speaker's corpora. The adversarial objective $\mathcal{L}_{rf}$
for this discriminator is formulated as follows:
\begin{equation}
\begin{aligned}
\mathcal{L}_{rf}(\mathcal{D}_{r})=
&\mathbb{E}_{y^{f}\sim{P(y^{f})}}[\mathcal{D}_{r}(y^{f})]\;+ \\
&\mathbb{E}_{y^{g}\sim{P(y^{g})}}[(1-\mathcal{D}_{r}(y^{g}))]
\end{aligned}
\end{equation}

\begin{equation}
\begin{aligned}
\mathcal{L}_{rf}(G)=
&\mathbb{E}_{y^{f}\sim{P(y^{f})}}[(1-\mathcal{D}_{r}(y^{f}))]
\end{aligned}
\end{equation}

Where $y^{g}$ is the ground-truth Mel-spectrograms and $y^{f}$ is the reconstructed Mel-spectrograms. 
The $\mathcal{L}_{rf}$ could improve the quality of generated spectrograms.

\textbf{Conversion discriminator} The conversion discriminator aims to distinguish whether a sequence of Mel-spectrograms has been converted by the acoustic model. The source speech of converted Mel-spectrograms comes from external unannotated corpora and is spoken by an external speaker, which leads to the speaker information contained in converted Mel-spectrograms differing from the target speaker due to the leaked residual source speaker information. Thus the converted Mel-spectrograms could be distinguished by the conversion discriminator. The adversarial objective $\mathcal{L}_{cvt}$
for this discriminator is formulated as follows:
\begin{equation}
\begin{aligned}
\mathcal{L}_{cvt}(\mathcal{D}_{c})=
&\mathbb{E}_{y^{c}\sim{P(y^{c})}}[\mathcal{D}_{c}(y^{c})]\;+\\
&\mathbb{E}_{y^{f}\sim{P(y^{f})}}[(1-\mathcal{D}_{c}(y^{f}))]\;+ \\
&\mathbb{E}_{y^{g}\sim{P(y^{g})}}[(1-\mathcal{D}_{c}(y^{g}))]
\end{aligned}
\end{equation}

\begin{equation}
\begin{aligned}
\mathcal{L}_{cvt}(G)=&\mathbb{E}_{y^{c}\sim{P(y^{c})}}[(1-\mathcal{D}_{c}(y^{c}))]
\end{aligned}
\end{equation}

Where $y^{c}$ is the converted Mel-spectrograms from external unannotated corpora, which have lower similarity because of the residual speaker information. 
We treated $y^{g}$ and $y^{f}$ as positive samples because there is no residual speaker information from other speakers. $\mathcal{L}_{cvt}$ could simulate a conversion step in the training stage, which could improve the similarity as well as the quality when the timbre of the input source speaker is different from the target speaker.

\textbf{Embedding discriminator} The Embedding discriminator aims to distinguish whether a sequence of content embedding is extracted from the external unannotated corpora, which are full of speaker information different from the target speaker. The adversarial objective $\mathcal{L}_{e}$
for this discriminator is formulated as follows:
\begin{equation}
\begin{aligned}
\mathcal{L}_{e}(\mathcal{D}_{e})=
&\mathbb{E}_{e^{o}\sim{P(e^{o})}}[\mathcal{D}_{e}(e^{o})]\;+\\
&\mathbb{E}_{e^{i}\sim{P(e^{i})}}[(1-\mathcal{D}_{e}(e^{i}))]\;
\end{aligned}
\end{equation}

\begin{equation}
\begin{aligned}
\mathcal{L}_{e}(G)=&\mathbb{E}_{e^{o}\sim{P(e^{o})}}[(1-\mathcal{D}_{e}(e^{o}))]
\end{aligned}
\end{equation}

Where $e^{i}$ is the content embedding extracted from speech belonging to the target speaker and $e^{o}$ is the content embedding extracted from speech belonging to an external speaker. $\mathcal{L}_{e}$ could force the generator to reduce the residual speaker information in content embedding and improve the similarity. Considering that both the $\mathcal{L}_{e}$ and $\mathcal{L}_{cvt}$ reduced the residual speaker information in different domains, we treated $\mathcal{L}_{e}$ and $\mathcal{L}_{cvt}$ as a similarity-adversarial loss:
\begin{equation}
\begin{aligned}
\mathcal{L}_{sim}(G)=&\mathcal{L}_{e}(G) + \mathcal{L}_{cvt}(G)\\
\mathcal{L}_{sim}(D)=&\mathcal{L}_{e}(\mathcal{D}_{e}) + \mathcal{L}_{cvt}(\mathcal{D}_{c})
\end{aligned}
\end{equation}

\subsection{Voice conversion loss} We computed the reconstruction loss between the ground truth and reconstructed Mel spectrograms. The reconstruction loss and the final objective loss functions are given as follows:

\begin{equation}
\begin{aligned}
\mathcal{L}_{rec} = \|y^{f}-y^{g}\|_{1}
\end{aligned}
\end{equation}

\begin{equation}
\begin{aligned}
\mathcal{L}(\mathcal{D})=&\mathcal{L}_{sim}(\mathcal{D}) + \mathcal{L}_{rf}(\mathcal{D}_{r})\\
\mathcal{L}(G)=&\mathcal{L}_{sim}(G) + \mathcal{L}_{rf}(G) + \mathcal{L}_{rec}
\end{aligned}
\end{equation}

\label{sec:method}

\section{Experiments}

\subsection{Training configurations}
All experiments were carried out in an any-to-one voice conversion setup and
almost all VC experiments were performed on CSMSC\cite{datakaber} which is a single-female-speaker Chinese speech corpus containing about 12 hours of recordings. The VC training data is augmented by changing the speaking rate from 0.8 to 1.2 to enhance the prosody diversity and 90\% of them were used as training data. In addition, we chose only the audios in Aishell-3\cite{shi2020aishell} as our external unannotated corpora for the training of $\mathcal{L}_{sim}$, which contain 218 different speakers and 85 hours in total.
For SSL representation, we extracted Hubert Soft unit, Hubert Raw feature from pretrained model\footnote{\url{https://github.com/bshall/hubert}}
published in \cite{van2022comparison} and 
ContentVec from pretrained model\footnote{\url{https://github.com/auspicious3000/contentvec}}
published in \cite{qian2022contentvec}. All the pretrained SSL models are trained on the Librispeech dataset\cite{panayotov2015librispeech}. We extract 256-dim Hubert Soft Unit as well as 768-dim ContentVec and Hubert Raw feature.
In training steps, the weight of $\mathcal{L}_{sim}$ is set to zero before 5000 steps. For comparison, we treated the proposed method without the embedding discriminator and the conversion discriminator as our baseline method.

To compare with the supervised recognition-synthesis-based VC method, we trained a VC model with the input of bottleneck features (BNFs) extracted from a supervised ASR. Following \cite{zhao2022disentangling}, we extract BNF from ASR trained with connectionist temporal classification loss (CTC-BNFs) due to its good disentangling characteristics, which lead to high similarity. Note that in the supervised VC method, the similarity is high even without $\mathcal{L}_{sim}$.
Demo pages is available now\footnote{\href{https://thuhcsi.github.io/icme2023-sslAdvVC/}{https://thuhcsi.github.io/icme2023-sslAdvVC/}}.

\begin{figure}[t]
  \centering
  \includegraphics[width=1.0\linewidth]{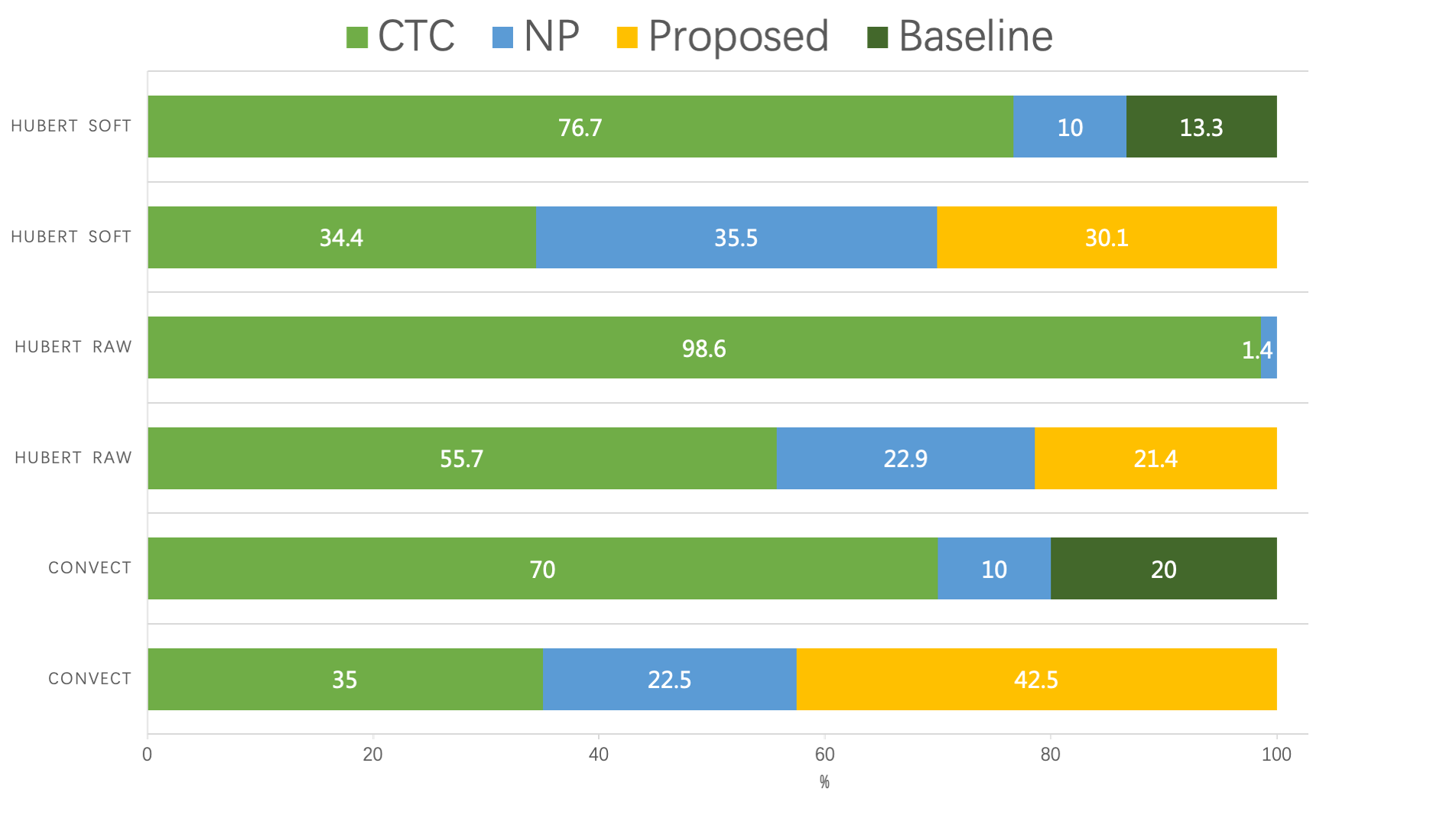}
  \caption{ABX test for similarity, NP means no preference.}
  \label{fig:ABX}
\end{figure}

\subsection{Subjective evaluation}
In order to compare with the supervised method, we conducted an ABX test to assess the timbre similarity of the CTC-BNFs method and the proposed VC models as well as the baseline VC models.10 source audios that come from different source speakers and scenes were used in all subjective experiments. Fig.\ref{fig:ABX} shows that the proposed methods could reach comparable similarity with CTC-BNFs, which has good disentangling characters, while baseline methods did much worse than CTC-BNFs. 

Mean opinion scores (MOS) tests were conducted to evaluate the converted waveform's prosody naturalness and timbre similarity. 
Each listener was asked to give a 5-point scale score with 1-point intervals. Table.\ref{tab:subj} shows that the proposed methods got high scores and significantly improved similarity compared to the baseline model while retaining high naturalness. Both subjective experiments show that: (i) The proposed method is applicable to other VC methods with different SSL representations as input. (ii) The proposed VC methods achieved comparable similarity with the supervised CTC-BNFs method in all chosen SSL representations.
Note that the CTC-BNFs method needs a massive amount of annotated corpora for ASR training, while the proposed method does not.

\begin{table}[h]
    \centering
    \small
	\caption{
	The MOS with 95\% confidence intervals in Naturalness and Similarity.
	}
	\label{tab:subj}
	\begin{tabular}{lcc}
		\toprule
		 &\small{Naturalness} 
		 &\small{Similarity} \\
		
        \hline
	\small{Hubert Soft} 
	&4.183(±0.193)
	&3.985(±0.212)  \\
		
        \small{w/o $\mathcal{L}_{sim}$ (Baseline)} 
        &4.267(±0.183)
        &3.584(±0.185)\\ 
        \hline
        \small{Hubert Raw} 
        &4.366(±0.249)
        &4.142(±0.278)\\
        \small{w/o $\mathcal{L}_{sim}$ (Baseline)} 
        &4.200(±0.316)
        &3.114(±0.216) \\ 
        \hline
        \small{ContentVec} 
        &3.771(±0.289)
        &4.030(±0.261)\\ 
        \small{w/o $\mathcal{L}_{sim}$ (Baseline)} 
        &4.086(±0.304)
        &3.550(±0.260)\\ 
        \hline
        \small{CTC BNFs} 
        &3.460(±0.235)
        &4.153(±0.197)\\ 
        
	\bottomrule
	\end{tabular}
\end{table}

\subsection{Objective evaluation}
Table.\ref{tab:Objct} shows the Mel-cepstral distortion (MCD), root-mean-square error (RMSE) in F0/energy and cosine-similarity (COS-SIM) calculated between speaker embedding. 
20 source audios from different speakers and scenes were
used in all objective experiments except the MCD experiment.
MCD and COS-SIM were used to measure how close the converted is to the target speech.
M2VoC parallel data \footnote{\href{http://challenge.ai.iqiyi.com/detail?raceId=5fb2688224954e0b48431fe0}{http://challenge.ai.iqiyi.com/detail?raceId=5fb2688224954e0b48431fe0}}(A male and a female speaker) were used for training and conversion in the MCD experiment.  
10 pairs of parallel audio were selected from two speakers in M2VoC as the test data. We calculated MCD between each converted audio and the parallel ground-truth audio coming from its target speaker, and the final MCD value is the average of respective experimental results in both male-to-female and female-to-male scenarios. 
For COS-SIM, speaker embeddings of the converted and the real target audio were extracted using a pretrained speaker verification model\cite{wang2022wespeaker}. A lower MCD value and higher COS-SIM indicate higher similarity. 
Energy and F0 RMSEs were used to measure the naturalness of the converted waveform. Both F0 and Energy sequences were performed min-max normalization before calculation. We calculated RMSE between the converted and source waveform because real speakers with fine-grained prosody and high naturalness uttered the source waveform 
, which means a lower RMSE value indicates higher naturalness and quality. 
The proposed methods achieve higher COS-SIM and MCD than the baseline models and are comparable to the supervised method with CTC-BNFs as input. On the other hand, the proposed methods achieve lower energy and F0 RMSEs than the CTC-BNFs method, which means more prosody information was preserved and higher naturalness. 

We also did an ablation study in Table.\ref{tab:Objct}. 
Comparing experiment w/o $\mathcal{L}_{sim}$ and w/o $\mathcal{L}_{e}$, we can find that the conversion discriminator significantly improved the timbre similarity. However, the prosody naturalness, as well as the quality, degrade.  
The comparison between the proposed method with and without $\mathcal{L}_{e}$ shows that if we introduce the embedding discriminator, the prosody information will be better preserved, improving the naturalness and the audio quality. In addition, the ablation of Real/Fake discriminator leads to the degradation in speech quality.

\begin{table}[h]
    \centering
    \small
	\caption{
 Objective experiments calculated between the waveforms generated by different systems.
 	Energy 
 	and F0 RMSEs were calculated between \textbf{source} and \textbf{converted} waveform while MCD and COS-SIM was calculated between 
 	\textbf{target} and \textbf{converted} waveform. 
	}
	\label{tab:Objct}
	\begin{tabular}{lcccc}
	\toprule
	\small{Model} &\small{MCD} &\small{COS-SIM} &\small{Energy RMSE} &\small{F0 RMSE}\\
        \hline

        \small{Hubert Soft} &\textbf{5.600} &\textbf{0.889} &0.156 &0.251 \\
        \small{w/o $\mathcal{L}_{r/f}$} &5.709 &0.877 &0.176 &0.296\\
        \small{w/o $\mathcal{L}_{e}$} &5.717 &0.880 &0.160 &0.294\\
	\small{w/o $\mathcal{L}_{sim}$} &5.813 &0.833 &0.119  &0.246 \\ 
        
        \hline
        \small{Hubert Raw} &\textbf{5.846} &\textbf{0.882} &0.152 &0.211 \\
        \small{w/o $\mathcal{L}_{r/f}$} &5.978 &0.880 &0.144 &0.371\\
        \small{w/o $\mathcal{L}_{e}$} &5.943 &0.883 &0.185 &0.330\\
        \small{w/o $\mathcal{L}_{sim}$} &5.970 &0.746  &0.103 &0.212 \\ 
        \hline
        \small{ContentVec} &5.739 &0.880 &0.159 &0.277 \\
        \small{w/o $\mathcal{L}_{r/f}$} &\textbf{5.664} &\textbf{0.891} &0.208 &0.338\\
        \small{w/o $\mathcal{L}_{e}$} &5.784 &0.883 &0.174 &0.311\\
        \small{w/o $\mathcal{L}_{sim}$} &5.797 &0.868  &0.136 &0.236 \\ 
        \hline
        \small{CTC BNFs} &5.476 &0.873 &0.173 &0.272 \\ 
	\bottomrule
        \end{tabular}
\end{table}

\subsection{Speaker information analysis}
\begin{figure}[t]
\begin{minipage}[b]{1.0\linewidth}
  \centering
  \includegraphics[width=1.0\linewidth]{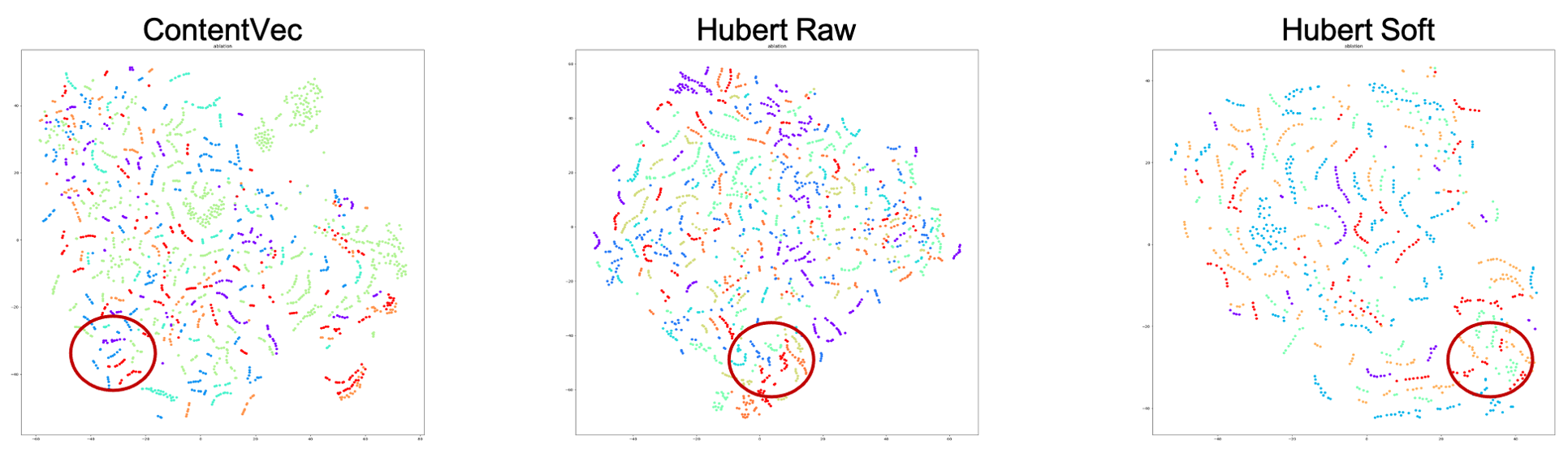}
  \centerline{(a) Without $\mathcal{L}_{sim}$}\medskip
\end{minipage}
\begin{minipage}[b]{1.0\linewidth}
  \centering
  \includegraphics[width=1.0\linewidth]{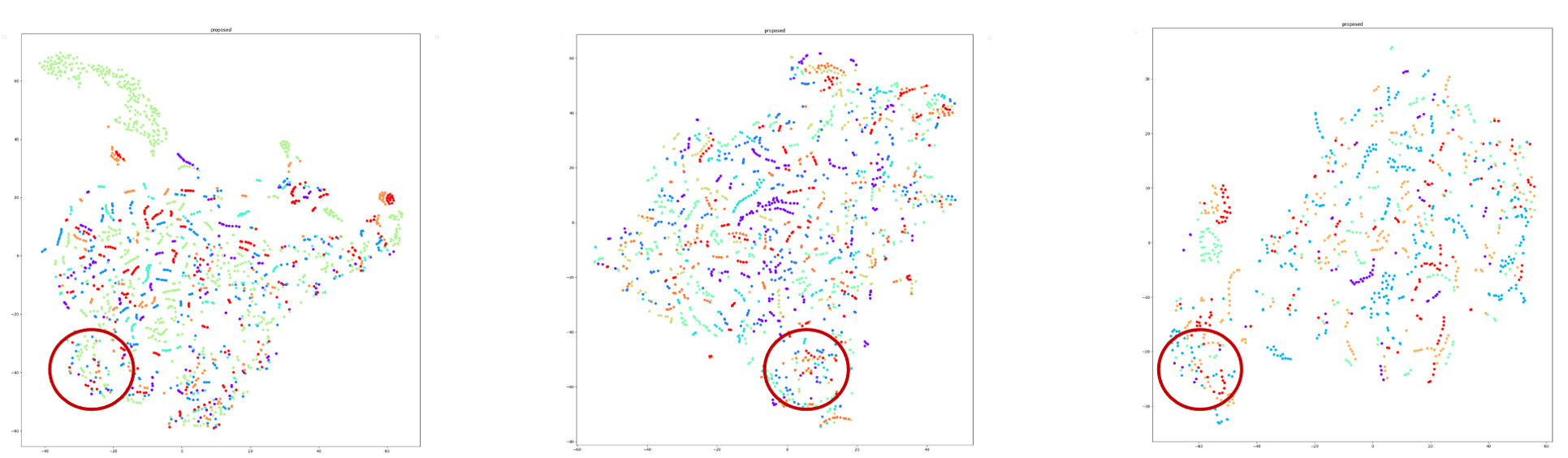}
  \centerline{(b) With $\mathcal{L}_{sim}$ }\medskip
\end{minipage}
\caption{T-SNE visualization in content embedding extracted from different SSL representations. The scatters in different colors belong to different speakers.}
\label{fig:visual_contemb}
\end{figure}
To show the residual speaker information leaked from the SSL representations, we visualized the extracted content embedding by T-SNE. As shown in Fig.\ref{fig:visual_contemb}, different colors correspond to different speakers. Fig.\ref{fig:visual_contemb} (a) shows that the content embedding extracted from the baseline method has been well-clustered, which means the content embeddings contain much pronunciation information. However, we can still find that in each cluster center, the point could be scattered by speakers, meaning there still is residual speaker information contained in it. In Fig.\ref{fig:visual_contemb} (b), we can find that the points are hardly scattered by speakers, demonstrating the proposed method with $\mathcal{L}_{sim}$ can reduce the leaked residual speaker information.

\label{sec:exp}

\section{Conclusion}
In this paper, we proposed a novel any-to-one recognition-synthesis-based SSL VC method with external unannotated corpora. We introduced similarity-adversarial loss to reduce the residual speaker information leaked from SSL representations, leading to superior performance than the baseline methods in similarity. The experiments show that the proposed VC methods can achieve comparable similarity and better naturalness than the supervised recognition-synthesis-based VC method, while no annotation was used in the training steps. In addition, we use the T-SNE method to demonstrate that the proposed methods could reduce the residual speaker information contained in content embedding. In the future, we will focus on any-to-many VC tasks and try to use more external unannotated corpora to reduce the naturalness loss.
\label{sec:conclusion}

\section*{Acknowledgment}

This work is supported by National Natural Science Foundation of China (62076144), National Social Science Foundation of China (13\&ZD189) and Shenzhen Science and Technology Program (WDZC20220816140515001, JCYJ20220818101014030).

\bibliographystyle{IEEEbib}
\bibliography{ref}

\begin{thebibliography}{10}
\providecommand{\url}[1]{#1}
\csname url@samestyle\endcsname
\providecommand{\newblock}{\relax}
\providecommand{\bibinfo}[2]{#2}
\providecommand{\BIBentrySTDinterwordspacing}{\spaceskip=0pt\relax}
\providecommand{\BIBentryALTinterwordstretchfactor}{4}
\providecommand{\BIBentryALTinterwordspacing}{\spaceskip=\fontdimen2\font plus
\BIBentryALTinterwordstretchfactor\fontdimen3\font minus
  \fontdimen4\font\relax}
\providecommand{\BIBforeignlanguage}[2]{{%
\expandafter\ifx\csname l@#1\endcsname\relax
\typeout{** WARNING: IEEEtranS.bst: No hyphenation pattern has been}%
\typeout{** loaded for the language `#1'. Using the pattern for}%
\typeout{** the default language instead.}%
\else
\language=\csname l@#1\endcsname
\fi
#2}}
\providecommand{\BIBdecl}{\relax}
\BIBdecl

\bibitem{baevski2020wav2vec}
A.~Baevski, Y.~Zhou, A.~Mohamed, and M.~Auli, ``wav2vec 2.0: A framework for
  self-supervised learning of speech representations,'' \emph{Advances in
  Neural Information Processing Systems}, vol.~33, pp. 12\,449--12\,460, 2020.

\bibitem{choi2021neural}
H.-S. Choi, J.~Lee, W.~Kim, J.~Lee, H.~Heo, and K.~Lee, ``Neural analysis and
  synthesis: Reconstructing speech from self-supervised representations,''
  \emph{Advances in Neural Information Processing Systems}, vol.~34, pp.
  16\,251--16\,265, 2021.

\bibitem{chou2019one}
J.-c. Chou, C.-c. Yeh, and H.-y. Lee, ``One-shot voice conversion by separating
  speaker and content representations with instance normalization,''
  \emph{arXiv preprint arXiv:1904.05742}, 2019.

\bibitem{datakaber}
Data-Baker, ``Chinese standard mandarin speech corpus,''
  \url{https://test.data-baker.com/data/index/TNtts/}.

\bibitem{gat2022speaker}
I.~Gat, H.~Aronowitz, W.~Zhu, E.~Morais, and R.~Hoory, ``Speaker normalization
  for self-supervised speech emotion recognition,'' in \emph{ICASSP 2022-2022
  IEEE International Conference on Acoustics, Speech and Signal Processing
  (ICASSP)}.\hskip 1em plus 0.5em minus 0.4em\relax IEEE, 2022, pp. 7342--7346.

\bibitem{hsu2021hubert}
W.-N. Hsu, B.~Bolte, Y.-H.~H. Tsai, K.~Lakhotia, R.~Salakhutdinov, and
  A.~Mohamed, ``Hubert: Self-supervised speech representation learning by
  masked prediction of hidden units,'' \emph{IEEE/ACM Transactions on Audio,
  Speech, and Language Processing}, vol.~29, pp. 3451--3460, 2021.

\bibitem{kong2020hifi}
J.~Kong, J.~Kim, and J.~Bae, ``Hifi-gan: Generative adversarial networks for
  efficient and high fidelity speech synthesis,'' \emph{Advances in Neural
  Information Processing Systems}, vol.~33, pp. 17\,022--17\,033, 2020.

\bibitem{liu2021fastsvc}
S.~Liu, Y.~Cao, N.~Hu, D.~Su, and H.~Meng, ``Fastsvc: Fast cross-domain singing
  voice conversion with feature-wise linear modulation,'' in \emph{2021 IEEE
  International Conference on Multimedia and Expo (ICME)}.\hskip 1em plus 0.5em
  minus 0.4em\relax IEEE, 2021, pp. 1--6.

\bibitem{panayotov2015librispeech}
V.~Panayotov, G.~Chen, D.~Povey, and S.~Khudanpur, ``Librispeech: an asr corpus
  based on public domain audio books,'' in \emph{2015 IEEE international
  conference on acoustics, speech and signal processing (ICASSP)}.\hskip 1em
  plus 0.5em minus 0.4em\relax IEEE, 2015, pp. 5206--5210.

\bibitem{qian2019autovc}
K.~Qian, Y.~Zhang, S.~Chang, X.~Yang, and M.~Hasegawa-Johnson, ``Autovc:
  Zero-shot voice style transfer with only autoencoder loss,'' in
  \emph{International Conference on Machine Learning}.\hskip 1em plus 0.5em
  minus 0.4em\relax PMLR, 2019, pp. 5210--5219.

\bibitem{qian2022contentvec}
K.~Qian, Y.~Zhang, H.~Gao, J.~Ni, C.-I. Lai, D.~Cox, M.~Hasegawa-Johnson, and
  S.~Chang, ``Contentvec: An improved self-supervised speech representation by
  disentangling speakers,'' in \emph{International Conference on Machine
  Learning}.\hskip 1em plus 0.5em minus 0.4em\relax PMLR, 2022, pp.
  18\,003--18\,017.

\bibitem{ren2019fastspeech}
Y.~Ren, Y.~Ruan, X.~Tan, T.~Qin, S.~Zhao, Z.~Zhao, and T.-Y. Liu, ``Fastspeech:
  Fast, robust and controllable text to speech,'' \emph{Advances in Neural
  Information Processing Systems}, vol.~32, 2019.

\bibitem{shi2020aishell}
Y.~Shi, H.~Bu, X.~Xu, S.~Zhang, and M.~Li, ``Aishell-3: A multi-speaker
  mandarin tts corpus and the baselines,'' \emph{arXiv preprint
  arXiv:2010.11567}, 2020.

\bibitem{sun2016phonetic}
L.~Sun, K.~Li, H.~Wang, S.~Kang, and H.~Meng, ``Phonetic posteriorgrams for
  many-to-one voice conversion without parallel data training,'' in \emph{2016
  IEEE International Conference on Multimedia and Expo (ICME)}.\hskip 1em plus
  0.5em minus 0.4em\relax IEEE, 2016, pp. 1--6.

\bibitem{van2022comparison}
B.~van Niekerk, M.-A. Carbonneau, J.~Za{\"\i}di, M.~Baas, H.~Seut{\'e}, and
  H.~Kamper, ``A comparison of discrete and soft speech units for improved
  voice conversion,'' in \emph{ICASSP 2022-2022 IEEE International Conference
  on Acoustics, Speech and Signal Processing (ICASSP)}.\hskip 1em plus 0.5em
  minus 0.4em\relax IEEE, 2022, pp. 6562--6566.

\bibitem{van2021analyzing}
B.~van Niekerk, L.~Nortje, M.~Baas, and H.~Kamper, ``Analyzing speaker
  information in self-supervised models to improve zero-resource speech
  processing,'' \emph{arXiv preprint arXiv:2108.00917}, 2021.

\bibitem{wang2021towards}
C.~Wang, Z.~Li, B.~Tang, X.~Yin, Y.~Wan, Y.~Yu, and Z.~Ma, ``Towards
  high-fidelity singing voice conversion with acoustic reference and
  contrastive predictive coding,'' \emph{arXiv preprint arXiv:2110.04754},
  2021.

\bibitem{wang2022wespeaker}
H.~Wang, C.~Liang, S.~Wang, Z.~Chen, B.~Zhang, X.~Xiang, Y.~Deng, and Y.~Qian,
  ``Wespeaker: A research and production oriented speaker embedding learning
  toolkit,'' \emph{arXiv preprint arXiv:2210.17016}, 2022.

\bibitem{wang2021adversarially}
J.~Wang, J.~Li, X.~Zhao, Z.~Wu, S.~Kang, and H.~Meng, ``Adversarially learning
  disentangled speech representations for robust multi-factor voice
  conversion,'' \emph{arXiv preprint arXiv:2102.00184}, 2021.

\bibitem{wang2022one}
Z.~Wang, Q.~Xie, T.~Li, H.~Du, L.~Xie, P.~Zhu, and M.~Bi, ``One-shot voice
  conversion for style transfer based on speaker adaptation,'' in \emph{ICASSP
  2022-2022 IEEE International Conference on Acoustics, Speech and Signal
  Processing (ICASSP)}.\hskip 1em plus 0.5em minus 0.4em\relax IEEE, 2022, pp.
  6792--6796.

\bibitem{zhao2022disentangling}
X.~Zhao, F.~Liu, C.~Song, Z.~Wu, S.~Kang, D.~Tuo, and H.~Meng, ``Disentangling
  content and fine-grained prosody information via hybrid asr bottleneck
  features for voice conversion,'' in \emph{ICASSP 2022-2022 IEEE International
  Conference on Acoustics, Speech and Signal Processing (ICASSP)}.\hskip 1em
  plus 0.5em minus 0.4em\relax IEEE, 2022, pp. 7022--7026.

\end{thebibliography}
\end{document}